# Spontaneous and stimulated emission tuning characteristics of a Josephson junction in a microcavity

Andrea T. Joseph, Robin Whiting and Roger Andrews

Department of Physics, University of the West Indies, St. Augustine, Trinidad and Tobago

randrews@fans.uwi.tt

We have investigated theoretically the tuning characteristics of a Josephson junction within a microcavity for one-photon spontaneous emission and for one-photon and two-photon stimulated emission.  For spontaneous emission, we have established the linear relationship between the magnetic induction and the voltage needed to tune the system to emit at resonant frequencies.  For stimulated emission, we have found an oscillatory dependence of the emission rate on the initial Cooper pair phase difference and the phase of the applied field.  Under specific conditions, we have also calculated the values of the applied radiation amplitude for the first few emission maxima of the system and for the first five junction-cavity resonances for each process.  Since the emission of photons can be controlled, it may be possible to use such a system to produce photons on demand.  Such sources will have applications in the fields of quantum cryptography, communications and computation.

*OCIS codes: (999.9999) Josephson junction; microcavity; spontaneous emission;  stimulated emission*

## 1. Introduction

Since the earlier half of the twentieth century, confined atoms have been studied both theoretically[1] and experimentally[2]. The interaction of these atoms or molecules with confined electromagnetic fields has always garnered a great deal of attention.[3,4]  The field of Cavity

Quantum Electrodynamics (CQED) deals with the study of such interactions in cavities.[5,6] Moreover, very often, these investigations involve the use of two-level atoms.[7,8] In 1970 and in 1976 respectively, Tilley,[9] and Rogovin and Scully[10] had presented theoretical studies illustrating the analogy between a Josephson junction and a two-level atom. The first experimental confirmation of their predictions came in 1999.[11] In this study, measurements were made of 150 GHz electromagnetic radiation emitted by three two-dimensional Josephson junction arrays. Each array was comprised of underdamped Nb / Al / AlO$_x$ / Nb Josephson tunnel junctions and each junction was coupled to a high-Q cell resonance. The arrays emitted coherently above a threshold that increased as the array size increased and the detected ac power exhibited an $N^2$ dependence, where $N$ represented the number of active junctions in the array. They also observed synchronization between incoming radiation and radiation emitted by the junctions. Even before this publication though, there had already been research carried out on the interaction of Josephson junctions with cavities.[12] Since 1999 other articles detailing analyses of Josephson junction-cavity systems have also been published.[13,14]

In this paper, Rogovin and Scully's model of the Josephson junction as a two-level atom is used to investigate specifically the spontaneous and stimulated emission of radiation from a Josephson junction within a microcavity. For spontaneous emission, the Hamiltonian employed uses the classical form of the pair current density for the junction in the presence of an applied static magnetic field and a DC voltage. For stimulated emission, with the junction placed in an external electromagnetic field and biased at a DC voltage, the form of the photon-assisted pair current density for the junction is used. The Josephson junction is assumed to be much smaller than the cavity length. The one-photon process is investigated for spontaneous emission and the one-photon and two-photon processes for stimulated emission. Relevant equations are presented to calculate the parameters necessary to achieve resonance for spontaneous emission and magnitudes of parameters needed to produce maximum count rates for stimulated emission are calculated.

## 2. Interaction Hamiltonian and field quantisation for a Josephson junction in a microcavity

The Josephson junction under consideration, is comprised of an insulating layer of thickness $l$ sandwiched between two superconducting layers and has been given a length $L$.

The resultant interaction Hamiltonian[15] describing the coupling of the electron pairs to the electromagnetic field is given as

$$\hat{H}_i = \frac{1}{c_o} \int d^3r \, \mathbf{j}(\mathbf{r}) \cdot \hat{\mathbf{A}}(\mathbf{r},t), \qquad (1)$$

where $c_o$ is the speed of the electromagnetic radiation in the insulating layer, $\mathbf{j}(\mathbf{r})$, the pair current density, and $\hat{\mathbf{A}}(\mathbf{r},t)$, the electromagnetic vector potential.

The Josephson junction is placed in the microcavity such that it is centred at the origin (Fig. 1). In a cavity, the positive frequency part of the quantised vector potential is defined as[16]

$$\hat{\mathbf{A}}^+(\mathbf{r},t) = \int d^3k \sum_j \left( \frac{\hbar}{16\pi^3 c_o k \varepsilon'} \right)^{1/2} \varepsilon(\mathbf{k},j) \left[ U_{\mathbf{k}j}(\mathbf{r}) \hat{a}_{\mathbf{k}j} + U'_{\mathbf{k}j}(\mathbf{r}) \hat{a}'_{\mathbf{k}j} \right] \exp(-ic_o k t). \qquad (2)$$

$\varepsilon(\mathbf{k},j)$ represents the unit vector for the transverse polarization directions where $j = 1, 2$ and $\varepsilon'$ is the permittivity of the medium of the cavity and insulating layer. The wave vectors can be either $\mathbf{k}_+$ or $\mathbf{k}_-$, where $\mathbf{k}_+$ is used for modes propagating to the right and $\mathbf{k}_-$, for those to the left. They are defined as[16]

$$\mathbf{k}_\pm = k(\sin\theta \cos\phi, \sin\theta \sin\phi, \pm\cos\theta), \qquad (3)$$

where $\theta$ $(0 < \theta < \pi/2)$ and $\phi$ $(0 < \phi < 2\pi)$ are the spherical angles. $U_{kj}(\mathbf{r})$ and $U'_{kj}(\mathbf{r})$ are the spatial mode functions of the cavity. $\hat{a}^+_{kj}$ and $\hat{a}_{kj}$ [$\hat{a}'^+_{kj}$ and $\hat{a}'_{kj}$] are the creation and destruction operators, respectively, for the mode with spatial function $U_{kj}(\mathbf{r})$ [$U'_{kj}(\mathbf{r})$].

The geometrical details of the Fabry-Perot microcavity used in this investigation are given in Fig. 2. The mirrors of the cavity, assumed to be ideal, with zero thickness and absorption, extend infinitely in the $x-y$ plane. The $z-$axis is perpendicular to the plane of the mirrors and the origins of the $x$, $y$ and $z$ axes are at the centre of the cavity.

The Josephson junction wave function, $|\psi_J\rangle$, is the direct product of the BCS wave functions for the upper and lower superconducting layers.[10] For the lower layer,

$$|\psi_n\rangle = |\mathbf{q}\uparrow, -\mathbf{q}\downarrow\rangle = \prod_{\mathbf{q}}(u_\mathbf{q} + v_\mathbf{q} C^+_{\mathbf{q}\uparrow} C^+_{-\mathbf{q}\downarrow})|0\rangle, \qquad (4)$$

and for the upper superconducting layer,

$$|\psi_m\rangle = |\mathbf{s}\uparrow, -\mathbf{s}\downarrow\rangle = \prod_{\mathbf{s}}(u_\mathbf{s} + v_\mathbf{s} C^+_{\mathbf{s}\uparrow} C^+_{-\mathbf{s}\downarrow})|0\rangle, \qquad (5)$$

where $(\mathbf{s}\uparrow,-\mathbf{s}\downarrow)$ [$(\mathbf{q}\uparrow,-\mathbf{q}\downarrow)$] represents the pair state of the electron pairs in the upper [lower] superconducting layer, and $C^+_{\mathbf{s}\uparrow}$ and $C^+_{-\mathbf{s}\downarrow}$ [$C^+_{\mathbf{q}\uparrow}$ and $C^+_{-\mathbf{q}\downarrow}$] are the creation operators for the electrons in the Bloch states $|\mathbf{s}\uparrow\rangle$ and $|-\mathbf{s}\downarrow\rangle$ [$|\mathbf{q}\uparrow\rangle$ and $|-\mathbf{q}\downarrow\rangle$] respectively. $u_\mathbf{s}$ [$v_\mathbf{s}$] is the probability amplitude that the pair state $(\mathbf{s}\uparrow,-\mathbf{s}\downarrow)$ is empty [filled]. $|\psi_J\rangle$ is given as $|\psi_n\rangle \times |\psi_m\rangle$, and the part of $|\psi_J\rangle$ that describes pair tunneling would include terms containing $C^+_{\mathbf{s}\uparrow} C^+_{-\mathbf{s}\downarrow}$ and $C^+_{\mathbf{q}\uparrow} C^+_{-\mathbf{q}\downarrow}$. For electrons with energy $\bar{\varepsilon}_\mathbf{q}$, where $\bar{\varepsilon}_\mathbf{q}$ is the electron energy relative to the Fermi energy, such that $-\hbar\omega_D < \bar{\varepsilon}_\mathbf{q} < \hbar\omega_D$, $u_\mathbf{q} = v_\mathbf{q} = \frac{1}{\sqrt{2}}$, where $\omega_D$ is the Debye frequency of the metal. Additionally, the dynamics of the pair tunneling under investigation are such that electrons are transferred across the junction with no change in their wave vectors. $|\psi_J\rangle$, can therefore be written as[10]

$$|\psi_J\rangle = \prod_\mathbf{s} \frac{1}{2}\left[C^+_{\mathbf{s},\uparrow,m} C^+_{-\mathbf{s},\downarrow,m} + C^+_{\mathbf{s},\uparrow,n} C^+_{-\mathbf{s},\downarrow,n}\right]|0\rangle. \qquad (6)$$

## 3. Spontaneous emission

The initial state of the junction is given in Eq. (6) and no photons are present in the cavity. The initial state of the system, $|I_{sp}\rangle$, can therefore be written as

$$|I_{sp}\rangle = |\psi_{J_I}, 0\rangle, \qquad (7)$$

where $|\psi_{J_I}\rangle$ represents the initial state of the Josephson junction. The final state, $|F_{sp}\rangle$, has the form

$$|F_{sp}\rangle = |\psi_{J_F}, 1\rangle, \qquad (8)$$

where $|1\rangle$ indicates the presence of a photon in the final state and $|\psi_{J_F}\rangle$ represents the final state of the junction.

The probability amplitude that the system, at time $t$ is in the state $|F\rangle$ is to first order in perturbation theory

$$-\frac{i}{\hbar}\int_{t_o}^{t}\langle F|\hat{H}_i|I\rangle dt' \qquad (9)$$

where $|I\rangle$ and $|F\rangle$ are the initial and final states respectively.

Substituting Eq. (1) into Eq. (9), and using the expressions for the initial $|I_{sp}\rangle$ and final $|F_{sp}\rangle$ states for spontaneous emission, we obtain

$$-\frac{i}{\hbar c_o}\int_{t_o}^{t}\int d^3r \left[\langle \psi_{J_F}|\mathbf{j}(\mathbf{r})|\psi_{J_I}\rangle \times \langle 1|\hat{\mathbf{A}}(\mathbf{r},t')|0\rangle\right]dt', \qquad (10)$$

To generate electromagnetic radiation from the junction through the process of spontaneous emission, a DC voltage, $V_o$, is established across the insulating layer and a small static magnetic field[17] is applied in the $y$-direction $(\mathbf{H}_o = 0, H_o, 0)$. The presence of the voltage causes the supercurrent to oscillate at a frequency of $2eV_o/\hbar$ [18] and electromagnetic waves are radiated from the junction. For a magnetic field directed along the $y$-direction, the solution of the vector potential in the oxide and superconducting layers are such that the dominant component lies in the direction of current flow. This conclusion results from the analysis carried out by Swihart,[19] and Eck, Scalapino and Taylor.[20] Due to the presence of the voltage and the magnetic field, the current component, $\langle \psi_{J_F}|\mathbf{j}(\mathbf{r})|\psi_{J_I}\rangle$, of Eq. (10) can be written as[10]

$$j_1 \sin(\omega_o t' - k_o z). \qquad (11)$$

$j_1$ is the pair current density amplitude and the Josephson frequency

$$\omega_o = 2eV_o/\hbar. \qquad (12)$$

$k_o$ is defined by the equation[10]

$$k_o = \frac{2e}{\hbar c_o}(2\lambda_L + l)H_o, \qquad (13)$$

and represents the effect of the externally applied magnetic field on the tunnelling of the Cooper pairs. $\lambda_L$ is the London penetration depth of the superconducting material and $c_o$ is given by

$$c_o = \left[\frac{l}{\varepsilon(2\lambda_L + l)}\right]^{1/2} c. \tag{14}$$

The field component of Eq. (10) is given as

$$\left(\frac{\hbar}{16\pi^3 c_o \varepsilon'}\right)^{1/2} \langle 1|\left(\int d^3k \sum_j k^{-\frac{1}{2}} \varepsilon(\mathbf{k},j)\left[U^*_{\mathbf{k}j}(z)\hat{a}^+_{\mathbf{k}j} + U'^*_{\mathbf{k}j}(z)\hat{a}'^+_{\mathbf{k}j}\right]\exp(ic_o kt')\right)|0\rangle, \tag{15}$$

where only the negative frequency part of the electromagnetic vector potential makes a non-zero contribution. The forms of the mode functions, $U_{\mathbf{k}j}(z)$ and $U'_{\mathbf{k}j}(z)$, in the region $-d/2 < z < d/2$ are as given below:[16]

$$U_{\mathbf{k}_+ j}(z) = \frac{t_{1j} \exp(ikz\cos\theta)}{D_j}, \tag{16}$$

$$U_{\mathbf{k}_- j}(z) = \frac{t_{1j} r_{2j} \exp(ik(d-z)\cos\theta)}{D_j}, \tag{17}$$

$$U'_{\mathbf{k}_+ j}(z) = \frac{t_{2j} r_{1j} \exp(ik(d+z)\cos\theta)}{D_j} \tag{18}$$

and

$$U'_{\mathbf{k}_- j}(z) = \frac{t_{2j} \exp(-ikz\cos\theta)}{D_j}, \tag{19}$$

where $r_{1j}$ and $t_{1j}$ [$r_{2j}$ and $t_{2j}$] are the complex reflection and transmission coefficients for mirror 1 [2] respectively and

$$D_j \equiv 1 - r_{1j} r_{2j} \exp 2ikd\cos\theta. \tag{20}$$

The amplitude is calculated along the z-axis, i.e. $\theta \cong 0°$. Another condition imposed is that mirror 1 is perfectly reflecting and mirror 2 highly reflecting, such that

$$r_{1j} = -1, \quad r_{2j} \to -1, \quad t_{1j} = 0, \quad t_{2j} \to 0. \tag{21}$$

Apart from constant factors, Eq. (15) is calculated to be

$$\int d^3k \left( k^{-\frac{1}{2}} \frac{\exp(ikz) - \exp(-ik(z+d))}{1 + r_{21}^* \exp(-2ikd)} \exp(ic_o k t') \right). \quad (22)$$

Eq. (11) and Eq. (22) are then substituted into Eq. (10) and after carrying out the spatial and time integrations, the steady-state amplitude works out to be proportional to

$$\frac{\omega_n^{3/2} \sin\left[\left(k_o - \frac{\omega_n}{c_o}\right)\frac{L}{2}\right]}{\left(1 + r_{21}^* \exp\left(-\frac{2i\omega_n d}{c_o}\right)\right)\left(\left(k_o - \frac{\omega_n}{c_o}\right)\frac{L}{2}\right)} \exp\left(-\frac{i\omega_n d}{c_o}\right). \quad (23)$$

To calculate resonant frequencies, $\omega_n$, that can be generated by the Josephson junction, the relation[21]

$$\omega_n = \frac{n\pi c_o}{L} \quad (24)$$

is used. By ensuring that the width, $d$, of the microcavity is an integral multiple of the length, $L$, of the Josephson junction, the resonant frequencies of the junction are resonant frequencies of the microcavity. From Eq. (23), there is resonance in the emission for

$$k_o \cong \frac{\omega_n}{c_o}. \quad (25)$$

This gives $B_n$ as

$$B_n = \frac{\hbar \omega_n \mu_o}{2e(2\lambda_L + l)}, \quad (26)$$

where $\mu_o$ is the permeability of free space. For each value of $\omega_n$ calculated, the corresponding voltage, $V_n$, is obtained from

$$V_n = \frac{\hbar \omega_n}{2e}. \quad (27)$$

## 4. Stimulated emission

### *4.1 The one-photon process*

To examine stimulated emission from the Josephson junction, a DC voltage, $V_o$, is maintained across the junction which is located in the microcavity in which there already exists an applied electromagnetic field. There is no static magnetic field present $(\mathbf{H}_o = 0)$ and $j(\mathbf{r})$ is now of the form[10]

$$j = J_1 \sin\left[\omega_o t + \frac{2e}{\hbar\omega} V_1 \sin(\omega t + \theta') + \phi_o\right], \tag{28}$$

which, expressed in terms of Bessel functions, is re-written as

$$j = J_1 \sum_{s=-\infty}^{\infty} J_s\left(\frac{2eV_1}{\hbar\omega}\right) \sin[(\omega_o + s\omega)t + \phi_o + s\theta']. \tag{29}$$

$J_1$ represents the current density amplitude of the Cooper pairs and $\omega_o$, the Josephson frequency. $V_1$ is the magnitude of the voltage induced by the applied field, $\omega$, the frequency of the external radiation, $\theta'$, the phase of the applied radiation and $\phi_o$, the initial Cooper pair phase difference. $J_s(2eV_1/\hbar\omega)$ represents the ordinary Bessel function of order $s$ and argument $2eV_1/\hbar\omega$.

To calculate the amplitude for stimulated emission, the Josephson junction-microcavity system is given an initial state such that one photon is present in the cavity and two photons are present in the final state. This is denoted respectively as

$$\begin{aligned}|I_{st}\rangle &= |\psi_{J_I}, 1\rangle, \\ |F_{st}\rangle &= |\psi_{J_F}, 2\rangle.\end{aligned} \tag{30}$$

$|I_{st}\rangle$ and $|F_{st}\rangle$ represent the initial and final states for stimulated emission.

The amplitude for stimulated emission is

$$-\frac{i}{\hbar c_o}\int_{t_o}^{t}\int d^3r \left[\langle\psi_{J_F}|j|\psi_{J_I}\rangle \times \langle 2|\hat{A}_x(z,t')|1\rangle\right]dt', \tag{31}$$

After carrying out the various integrations, Eq. (31) works out to be proportional to

$$\frac{\omega_n^{1/2}\sin\left(\frac{\omega_n L}{2c_o}\right)\left(1-\exp\left(-i\frac{\omega_n d}{c_o}\right)\right)\exp(-i\theta')}{\left(1+r_{21}^*\exp\left(-i\frac{2\omega_n d}{c_o}\right)\right)}\left(J_{-2}\left(\frac{2eV_1}{\hbar\omega_n}\right)\exp(i\Delta\varphi)-J_0\left(\frac{2eV_1}{\hbar\omega_n}\right)\exp(-i\Delta\varphi)\right)$$

(32)

where $\omega_n$ represents resonant frequencies of the junction-cavity system. Here, again, $\omega_n$ is given by Eq. (24) and we are considering a situation where the resonant frequencies of the bare junction coincide with the resonant frequencies of the microcavity. $\Delta\varphi$ is given as

$$\Delta\varphi = \phi_o - \theta'. \tag{33}$$

To obtain Eq. (32), the applied microwave frequency is taken to be equal to one of the resonant frequencies of the system since we are considering one photon processes. For one photon emission, the current in Eq. (29) is such that $s=0$ and $s=-2$.

Under the conditions of resonance, the one photon stimulated emission amplitude is proportional to

$$J_{-2}\left(\frac{2eV_1}{\hbar\omega_n}\right)\exp(i\Delta\varphi)-J_0\left(\frac{2eV_1}{\hbar\omega_n}\right)\exp(-i\Delta\varphi) \tag{34}$$

where $0 \leq \Delta\varphi \leq 2\pi$.

### *4.2 The two-photon process*

For this process, the Josephson junction-microcavity system is in an initial state such that one photon is present in the cavity and three photons are present in the final state. The initial and final states, therefore, are denoted respectively as

$$|I_{st'}\rangle = |\psi_{J_I}, 1\rangle,$$
$$|F_{st'}\rangle = |\psi_{J_F}, 3\rangle. \quad (35)$$

To second order in perturbation theory, the probability amplitude that the system at time $t$ is in the state $|F_{st'}\rangle$ is[22]

$$\langle F_{st'}| \frac{1}{2}\left(-\frac{i}{\hbar}\int_{t_o}^{t} \hat{H}_i(t')dt'\right)^2 |I_{st'}\rangle \quad (36)$$

Ignoring constants, the amplitude for stimulated emission of the two-photon process is therefore calculated from

$$\int_{t_o}^{t} dt' \int_{t}^{t} dt'' \int d^3r' \int d^3r'' \left[\langle \psi_{J_F}|j(t')j(t'')|\psi_{J_I}\rangle \times \langle 3|\hat{A}_x^-(z',t')\hat{A}_x^-(z'',t'')|1\rangle\right]. \quad (37)$$

The same conditions hold as for the one-photon process. However,

$$\omega_o = 2\omega = 2\omega'. \quad (38)$$

($\omega_o$ is the Josephson frequency, $\omega$ is the frequency of the external radiation and $\omega'$ is the frequency of the radiation generated by the junction.) For two-photon emission, the current in Eq. (29) is such that $s = -1$ and $s = -3$. The amplitude for stimulated two-photon emission therefore works out to be proportional to

$$\left(\frac{\omega_n^{1/2} \sin\left(\frac{\omega_n L}{4c_o}\right)\left(1-\exp\left(-i\frac{\omega_n d}{2c_o}\right)\right)}{\left(1+r_{21}^* \exp\left(-i\frac{\omega_n d}{c_o}\right)\right)}\right)^2 \exp(-2i\theta')$$

$$\times \left(\left(J_{-3}\left(\frac{4eV_1}{\hbar\omega_n}\right)\right)^2 \exp(i2\Delta\varphi')+\left(J_{-1}\left(\frac{4eV_1}{\hbar\omega_n}\right)\right)^2 \exp(-i2\Delta\varphi')\right. \qquad (39)$$

$$\left. -2J_{-3}\left(\frac{4eV_1}{\hbar\omega_n}\right)J_{-1}\left(\frac{4eV_1}{\hbar\omega_n}\right)\right).$$

$\Delta\varphi'$ is given as

$$\Delta\varphi' = \phi_o - 2\theta'. \qquad (40)$$

At resonance, the two-photon stimulated emission amplitude is proportional to

$$\left(J_{-3}\left(\frac{4eV_1}{\hbar\omega_n}\right)\right)^2 \exp(i2\Delta\varphi')+\left(J_{-1}\left(\frac{4eV_1}{\hbar\omega_n}\right)\right)^2 \exp(-i2\Delta\varphi')-2J_{-3}\left(\frac{4eV_1}{\hbar\omega_n}\right)J_{-1}\left(\frac{4eV_1}{\hbar\omega_n}\right), \qquad (41)$$

where $0 \leq \Delta\varphi' \leq 2\pi$.

## 5. Results

We analyse a junction-cavity system for which values of $\varepsilon$, $l$, $L$, $\lambda_L$ and $d$ are taken as 10, 2 nm, 10 μm, 0.05 μm and 150 μm, respectively. From Eqs. (26) and (27), predicted numerical values of the magnetic induction, $B_n$, and the voltage, $V_n$, may be calculated for the first five resonant frequencies of the junction-cavity system. These values of $B_n$ and $V_n$ can be used to tune the junction for spontaneous emission of radiation at these specific frequencies.

In the case of stimulated emission, the amplitude in Eq. (34) is used to determine the count rate for the one-photon process and the amplitude in Eq. (41) is used for the two-photon process. In order to find the conditions of resonance, we need to determine the values of the argument of the Bessel functions $(X' = 2eV_1/\hbar\omega_n)$ such that the count rate is a maximum for a given $\Delta\varphi$. For the one-photon process, Fig. 3 (a) is a plot of the modulus square of the amplitude against $X$ for $\Delta\varphi = 0$. The count rate exhibits a decaying oscillatory dependence on $X$. The first three values that give a resonance in the count rate are $X = 3.518$, $X = 6.866$ and $X = 10.073$. These values of $X$ are used to find $V_1$ for a particular resonant frequency of the system. Fig. 3 (b) gives the numerical values of $V_1$ and $\omega_n$ for the three values of $X$. These predicted values of $V_1$ and $\omega_n$ can be used to tune the junction-cavity system. Fig. 4 (a) is a similar plot to 3 (a) for the two-photon process. This graph displays four values of $X'$ for which the count rate is a maximum – $X'_1 = 1.558$, $X'_2 = 4.888$, $X'_3 = 8.289$ and $X'_4 = 11.530$. While there is also a decaying oscillatory dependence as for the one-photon process, this begins only after the second position of maximum count rate. Fig. 4 (b) gives the numerical values of $V_1$ that correspond to the first five resonant frequencies of the junction-cavity system. The voltages $V_n$ required to achieve the resonant frequencies for both the one-photon and two-photon processes may be obtained using Eq. (27). Figs. 5 (a) and (b) show the dependence of the stimulated emission count rate on $\Delta\varphi$ for the one-photon process and on $\Delta\varphi'$ for the two-photon process. There is clearly an oscillatory dependence of the count rate on both parameters $\Delta\varphi$ and $\Delta\varphi'$.

In Al-Saidi and Stroud's study[23] of a single small underdamped Josephson junction in a large-Q resonant electromagnetic cavity, they employed the full Hamiltonian for the system whereas here, we are interested only in the interaction part of the Hamiltonian. From their analysis they found that for special values of the gate voltage, there was strong interaction between the junction and the resonant photon mode. In this paper it was found that resonance in the emission depended on the matching of the voltages and magnetic inductions (Eqs. (25) to (27)) for spontaneous emission, and for stimulated emission, on the relation of the voltage induced by the applied field to the potential difference across the junction (which can be determined from the argument of the Bessel function $X'$) and on the relationship between the phase of the applied field and the initial Cooper pair phase difference (Eq. (40)).

## 6. Conclusion

We have presented a model that describes the emission characteristics of a Josephson junction in a microcavity. For spontaneous emission, we have shown that the magnetic induction, $B_n$, and voltage, $V_n$, across the junction must be properly matched to achieve resonance in the cavity. For emission in the microwave region, typical values are of the order of milliteslas for $B_n$, and millivolts for $V_n$. For the stimulated emission processes, we calculated the amplitude of the applied radiation, $V_1$, necessary to produce maximum count rates for the first five resonant frequencies of the system. We have also shown that to achieve these maximum count rates for the one-photon process, the difference, $\Delta\varphi$, between the initial Cooper pair phase difference and the phase of the applied field must be an integral multiple of $\pi$, while for the two-photon process, the difference, $\Delta\varphi'$, between the initial Cooper pair phase difference and twice the phase of the applied field must be an odd number multiple of $\pi/2$. The ability to control the Josephson junction-cavity system to emit particular numbers of photons, to emit at particular frequencies, and to produce specific count rates as required certainly makes it a versatile source of photons. In particular, single-photon sources will find useful applications in quantum cryptography and information processing.

**List of Figure Captions**

Fig. 1. Schematic of a tunnel or S-I-S (S – superconductor, I – insulator) Josephson junction in a microcavity.

Fig. 2. Geometry of the Fabry-Perot microcavity. 1 and 2 represent the left and right mirrors, respectively, of the microcavity.

Fig. 3. Graphs illustrating the numerical values for the one-photon process of (a) X for the first three positions of maximum count rate for $\Delta\varphi = 0$, and (b) $V_1$ for the first five cavity resonances of the system for $X_1 = 3.518$, $X_2 = 6.866$ and $X_3 = 10.073$ and $\Delta\varphi = 0$.

Fig. 4. Graphs illustrating the numerical values for the two-photon process of (a) X for the first four positions of maximum count rate for $\Delta\varphi' = 0$, and (b) $V_1$ for the first five cavity resonances of the system for $X_1' = 1.558$, $X_2' = 4.888$, $X_3' = 8.289$ and $X_4' = 11.530$ and $\Delta\varphi' = 0$.

Fig. 5 Graph illustrating the variation of the count rate with (a) $\Delta\varphi$, where $\Delta\varphi = \phi_o - \theta'$, for the one-photon process and (b) $\Delta\varphi'$, where $\Delta\varphi' = \phi_o - 2\theta'$, for the two-photon process.

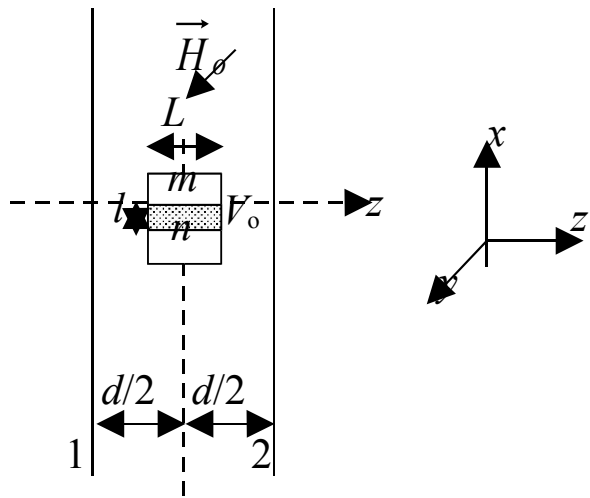

**Fig. 1**

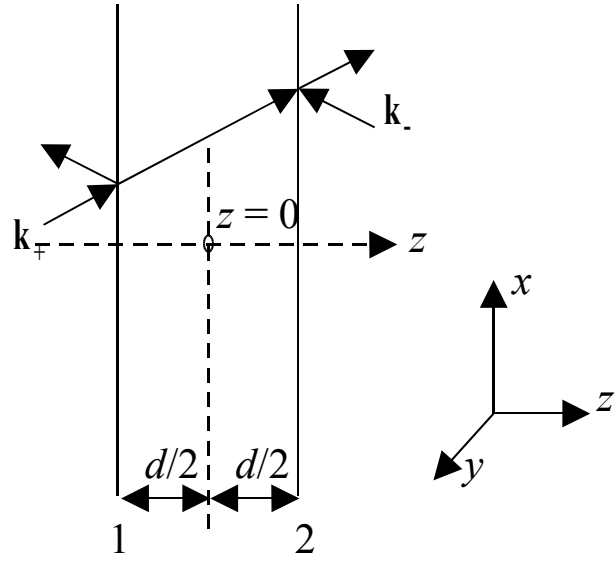

**Fig. 2**

24.

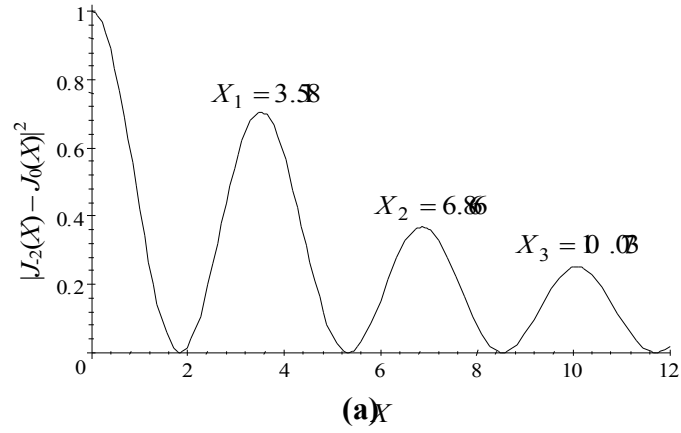

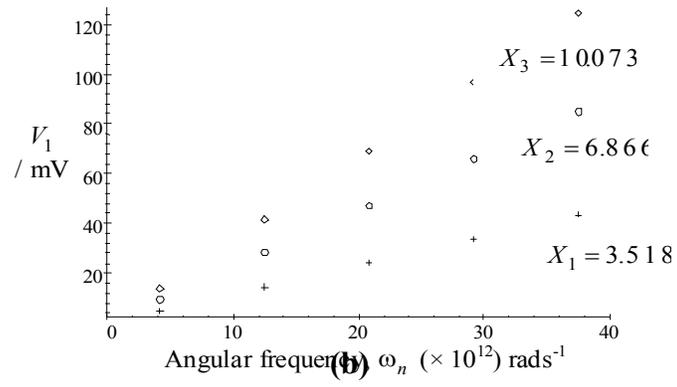

**Fig. 3**

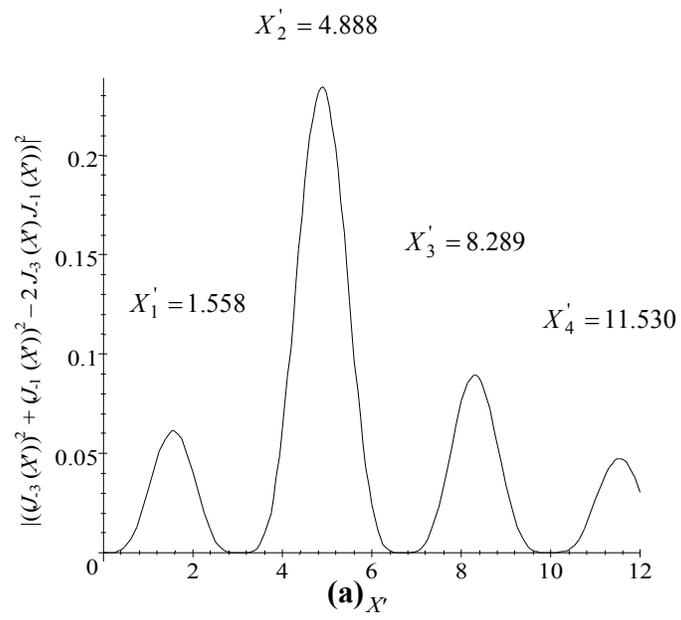

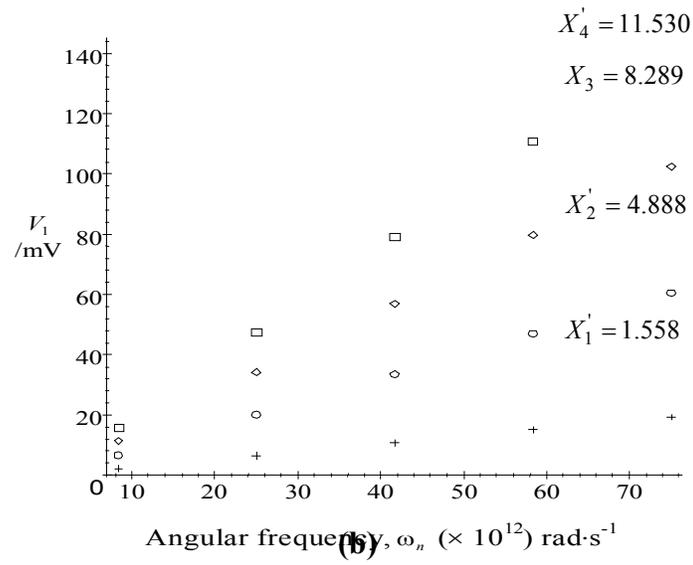

**Fig. 4**

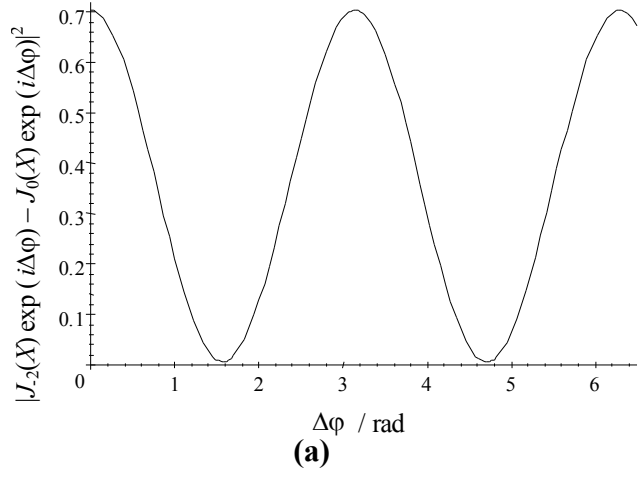

(a)

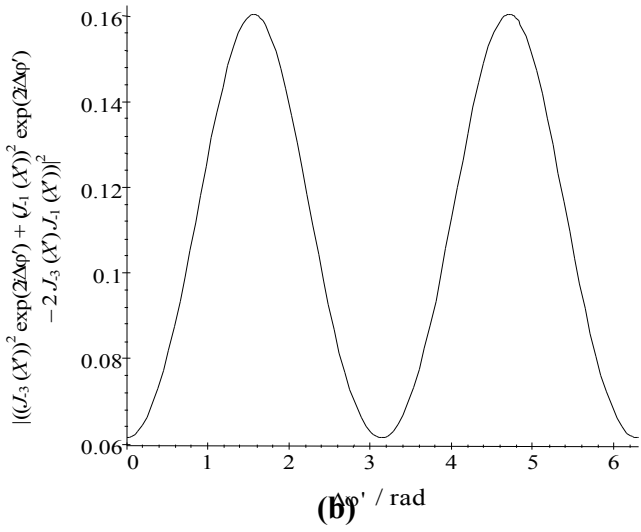

(b)

**Fig. 5**